A Silicon Valley Love Triangle: Hiring Algorithms, Pseudo-Science, and the Quest for Auditability

MONA SLOANE[1]
New York University and Tübingen University, ms11521@nyu.edu

EMANUEL MOSS[2]
Cornell Tech and Data & Society Research Institute, emanuelmoss@cornell.edu

Rumman Chowdhury
Director of ML Ethics, Transparency and Accountability (META) at Twitter, rchowdhury@gmail.com

Corresponding Author: Mona Sloane (ms11521@nyu.edu)
Lead Contact: Emanuel Moss (emanuelmoss@cornell.edu)

SUMMARY

In this paper, we develop a matrix for auditing algorithmic decision-making systems (ADS) used in the hiring domain. The tool is a socio-technical assessment of hiring ADS that is aimed at surfacing the underlying assumptions that justify the use of an algorithmic tool and the forms of knowledge or insight they purport to produce. These underlying assumptions, it is argued, are crucial for assessing not only whether an ADS works "as intended", but also whether the intentions with which the tool was designed are well-founded. Throughout, we contextualize the use of the matrix within current and proposed regulatory regimes, and within emerging hiring practices that incorporate algorithmic technologies. We suggest using a matrix to expose underlying assumptions rooted in pseudoscientific essentialized understandings of human nature and capability, and to critically investigate emerging auditing standards and practices that fail to address these assumptions.

CCS CONCEPTS • Social and Professional Topics

**Additional Keywords and Phrases:** Audit, Accountability, Hiring, Automated Decision Making Systems

---

[1] corresponding author
[2] lead contact

# 1. Introduction

Prospective job applicants find their interactions with future workplaces increasingly algorithmically mediated through automated decision making systems (ADS) that intervene throughout the hiring process[1,2]. These systems have long been used to scan hiring platforms for likely candidates, review resumes to filter in applicants who meet experience and credential criteria, rank applicants, and manage the hiring workflow for hiring managers. More recently, a suite of algorithmic applicant screening technologies are being integrated into the hiring process. These screening technologies variously evaluate applicants by assessing their aptitude for a role through online game playing, analyzing their speech and/or mannerisms to predict on-the-job performance, or by analyzing Meyers-Briggs styled "personality assessment" questionnaires.[3] These systems typically use NLP, computer vision, or supervised machine learning techniques that claim to predict job performance based on intonation, written text, micro-expressions, or game performance. While reliable metrics to evaluate the likely success of applicants remain elusive,[4] applicant screening and pre-interview evaluation tools are nevertheless being promoted as if they provide stable, reliable, objective, and fair insights into applicants' suitability for roles without regard for the validity of the evaluative constructs being deployed.[5,6] Most worryingly, these constructs are often grounded in pseudo-scientific practices[7,8] that recall the dark, eugenicist histories of physiognomy.[9,10]

Even though there is mounting evidence that such systems harbor bias across demographic categories,[11] algorithmic and bureaucratic opacity[12-14] have led to slow responses from regulators. Indeed, many of these systems promise more equitable outcomes when compared to the purported biases of human decision-makers. While recent work has examined how hiring managers operationalize concepts like "fairness" through their interactions with these algorithmic systems,[15] more research is needed to trace the ways claims about the revelatory capacity of hiring ADSs—the claims that they reveal the true potential of job candidates— are shaping the hiring ecosystem. This article contends that audits and assessments of hiring ADSs cannot be limited merely to the degree to which they promote demographic parity within the hiring process, but rather must also contend with claims that such ADSs can reveal aptitude, future performance, and cultural fit in order to promote equity and accountability across the entire hiring ecosystem.

# 2. Auditing efforts

In response to investigations that demonstrate a broad range of inequities produced by the integration of ADSs in society,[16-18] a new "algorithmic auditing" industry is blossoming to interrogate and document the potentially discriminatory effects of ADSs.[19-21] Actors within this nascent discipline have been left to their own devices in laying out frameworks and standards to satisfy the demand for audit practitioners coming from industry. This has resulted in what we argue here is an inappropriately constrained form of audit focused on metrics for algorithmic performance.[22] Brown et al.[19], for example, focus on five categories of metrics that robustly characterize the behavior of an algorithmic system, but such metrics are limited in assessing the system within a broader historical or socio-technical context. Algorithmic audits may demonstrate the veracity of a range of claims about an ADS being audited, but their scope is often limited to a narrow range of concerns set in advance by the ADS developer, and most often limited to auditing the performance of ADSs with respect to bias across legally protected categories like race and gender, e.g. Wilson et al. 2021.[23]

Recently, there have been significant regulatory proposals that call for algorithmic audits in the US[24,25] and EU.[26] Such proposals do not provide clear regulatory definitions of algorithmic audits; proposed bills such as New York City's bill A1894-2020,[25] which mandates annual "bias audits" of hiring ADSs, and the proposed Algorithmic Accountability Act in the US Congress[24] which calls for algorithmic impact assessments, do not dictate what precisely they should entail. This is also true of the Regulation on European Approach for Artificial Intelligence, which calls for audits and "conformity assessments" without providing instruction on what these practices ought to entail.[26] This leaves a regulatory grey area upon which the relevance of such bills hinge. That is, depending on the depth, breadth, and focus of audits as they come into practice through legislation, a bill like A1894-2020 could either enact unprecedented, meaningful steps towards enabling transparency and accountability, or create a bureaucratic shield for unscrupulous companies to hide behind by accountability-washing their ADS products. In this later case, toothless legislation could further deprive impacted individuals of their agency within an already opaque, burdensome, and inequitable hiring ecosystem. Developing robust forms of audit and assessment, and firmly establishing them within existing industry practices, is therefore a crucial step toward ensuring that any forthcoming regulatory and legislative interventions can create meaningful accountability for ADSs.

Currently, there are much-needed ongoing debates about whether algorithmic accountability ought to focus on industry- or sector-specific auditing techniques or on general, all-purpose algorithmic audits.[21,27-29] Additionally, there are debates about what the proper unit of analysis for algorithmic accountability ought to be. These debates ask if audits and assessments ought to evaluate a trained algorithmic model, a trained model and the data it was trained on, or a product that may contain multiple interlinked algorithmic models (Bandy 2021, Brown et al. 2021, Koshiyama et al. 2021, Sandvig et al. 2014).[19-21,30] These broader questions about the scope and target of audits are important, but within the hiring industry, there are specific challenges thought ought to be considered, and doing so can also shed light on how to address considerations that must be made as part of a regulatory approach to algorithmic accountability.

## 2.1 Auditing Hiring ADSs

Within the hiring domain, companies who use ADSs seldom use one single model for hiring Instead they, use a suite of ADSs that feed decisions into each other. A job seeker applying for a single open position may encounter different ADSs that recruit them to apply based on their social media profile, that parse their application looking for qualifications and credentials, that subject them to aptitude or personality tests, and that compile a synoptic profile for the hiring manager that will ultimately decide whether to hire them. Therefore, individual ADSs ought to be assessed not as if they are a stand-alone, independent tool, but rather with a consideration of the position they occupy within a linear series of interdependent models that feed one into the other. Put differently, any assessment may treat a tool as its unit of analysis, but cannot be assessed as if it operates free from its context of use. In the hiring domain, the outputs of one ADSs are often used as the inputs of another,[31] and any one of those tools should be evaluated with those dependencies in mind.

From the perspective of job applicants and hiring managers who wish to understand how the tools they use affect their work, the macroscopic process is only as transparent and accountable as its weakest, or most intransparent and unaccountable, ADS. Furthermore, ADS components that are downstream of other components are limited to the outputs of components that are computationally prior to them. Therefore, a tool used in a later stage of hiring may inherit the demographic biases of a tool used at an earlier stage. As an example, assume an organization has a simple hiring pipeline comprised of an ADS that identifies and reaches out to potential candidates encouraging them to apply and an ADS that ranks resumes using natural language processing to evaluate how well that resume fits a job description. If that company audits the resume parsing tool to ensure that it operates in accordance with a corporate value that prioritizes improving gender diversity, that audit would focus on the biases introduced by language models and the potential for gender-based language biases affecting the resume parsing process. However, even if this ADS were assessed and corrected for language biases, it is only as unbiased as the model before it—an unrepresentative sample of the potential applicant pool might have been produced by a biased recruiting tool. If the outreach ADS discriminates against, say, non-male candidates, then the downstream model can only perform to the ceiling set by the prior, biased, model.

Feeding the outputs of the resume parser into a third model, for example an "emotion detection" ADS that determines candidate trustworthiness by analyzing a video interview, an additional dimension of complexity is introduced in that the "biases" that come into consideration exist not as a function of data or model choice, but in the epistemological roots of the system. By "epistemological roots", we are referring to the claims to knowledge that the system is making—specifically that there is a way to "know" the interior emotional state of a subject based on externally discernable attributes like facial expressions, pupil dilations, or other physiological characteristics. Such claims have largely failed to demonstrate scientific validity, have not been replicated experimentally,[32,33] do not support the additional claims made by vendors that they are useful in predicting on-the-job performance,[34] and most troublingly replicate pseudo-scientific and flawed research that posits imagined links between biology and trustworthiness.[9,10] Such a system cannot be assessed as to whether it operates "as intended," if the intention is based on invalid claims about the relationship between appearance and performance. Audits and other investigations that are limited to disparate impact, gender distributions in the data, and the like cannot account for or correct this later tranche of problems.

## 3. The Socio-Technical Matrix for Assessing ADS

Against that backdrop, it is clear that current approaches to audit are inadequate at addressing the entire set of claims made by ADSs if they do not also consider the intentions and claims to knowledge that underlie their operation. If regulation and legislation are to be effective in producing accountability through mandating or recommending audit or assessment, methods for conducting such audits and assessments need to include new ways of framing and understanding how technological systems are encountered in the course of social life. For job applicants' encounters with technological systems, particularly when those systems build on pseudo-scientific theories, the stakes of inadequate audits and assessments are particularly instructive. To better resolve these takes, and to aid in ongoing efforts to build effective ways of framing and understanding the encounter between job applicants and ADSs, we propose an evaluative matrix for developing a holistic view of hiring ADSs by combining information on its context, its goal, its data, its function, its assumption, and epistemological roots. This matrix can serve as a template for auditors and other assessors, but can also be used to support new mechanisms for literacy, accountability, and oversight of ADSs for workers, researchers, policymakers and practitioners alike. Whereas other model documentation practices like "datasheets for datasets"[35] or "model cards for model reporting"[36] are intended to be compiled by developers, this matrix is intended to also be used by those outside the development and documentation process, to assess already-extant algorithmic products.

### 3.1 How the Socio-Technical Matrix Works

The Socio-Technical Matrix is a research tool that can serve as a basis for developing holistic socio-technical assessment and audit methods for hiring ADS. The matrix can be used by ADS developers, hiring managers interrogating the ADS tools they use throughout their hiring pipeline, and by the general public (including advocates for job-seekers and job-seekers themselves). In order to use the matrix, information on the hiring ADS needs to be collected. In contrast to high-level guidance offered for the self-assessment of algorithmic systems,[37] which ask developers to affirm whether or not they contemplated various considerations relevant to the trustworthy development of algorithmic technologies, the matrix offers prompting questions intended to produce documentary evidence of the answers to these questions, and suggests methods for doing so. Access to this information will vary depending on who is using the matrix, as companies using such systems are not obliged to disclose to candidates or the public that they are using hiring ADSs, or which hiring ADSs they are using. However, information on these hiring ADSs can be found on the Internet as vendors advertise their products and services via case studies, or in federal trademark filings.[14]

Some vendors offer a single hiring ADS to be used for a narrow purpose (such as the use of a resume parser to narrow down prospects), while other companies offer a suite of hiring tools. For the purposes of this paper, the unit of analysis for the matrix is individual ADSs that may be used at specific stages of the hiring process, not the entire process or the company using ADSs (although company-specific uses of the algorithm are relevant for filling out the matrix).

The matrix is comprised of seven elements which constitute a description of a hiring system and which need to be assembled in order to assess its claims.

The matrix prompts an auditor or assessor to compile relevant **information** about each element, and suggests useful **questions and methods** for obtaining that information (see Table 1). The seven elements are:

The **Hiring ADS** and **Funnel Stage** identify what the hiring ADS is (eg. Hubert.ai, ZipRecruiter) and how it is intended to be used within the "hiring funnel". The hiring funnel is a heuristic developed by Bogen and Reike[31] that segments the hiring process into successive stages in which ADSs are variously employed. These stages are the **sourcing** of potential job candidates, the **screening** of candidates to assess their general appropriateness for an open role, the **interviewing** of applicants to gauge their individual suitability for position, and the **selecting** of applicants from among a small pool of suitable candidates.

The **Goal** of the hiring ADS should clearly state what it aims to do (eg. "Filter the top 1% of applicants while maintaining the diversity of the applicant pool"). These three elements can be derived from sales copy, but should be supplemented by interviews with developers and hiring managers who purchase and use the hiring ADS.

Automated hiring ADSs use **Data**. Some hiring ADSs, particularly those that use machine learning or make claims about using artificial intelligence, use data other than that provided by a job applicant to sort, rank, filter, and predict performance for applicants. The matrix should help identify the data provided by applicants is processed (such as resumes, facial images, voice recordings, chatbot histories, gameplay).

The **Function** of a hiring ADS is a plain-language description of how it processes data to make its claims (eg. "compares resumes of previously successful employees to current applicants to predict future success"). Information pertaining to the function of an hiring ADS, how it processes data, and access to the hiring ADS itself should be procured through arrangements with developers and the hiring managers that configure and operate that hiring ADS. The model, or the hiring ADS itself, can subsequently be inspected as a part of an audit or impact assessment by examining the machine learning model that pursues this function in the context of training data, parameter settings, performance characteristics, and its integration into the hiring funnel.

The **Assumptions** that undergird a hiring ADS should capture the logic by which a hiring ADS is seen as useful, and can sometimes be derived from sales copy, but should a more thorough understanding of a hiring ADS's assumptions can be gathered from interviewing developers who create a hiring ADS and hiring managers who use a hiring ADS. These assumptions take the form of: "This hiring ADS works by comparing the resumes of successful employees to new applicants, because successful hires have proven that the attributes documented in their resumes are good predictors of future success."

Establishing the **Epistemological Roots** of a hiring ADS requires archival and/or ethnographic research to outline how a hiring ADS is understood to produce useful knowledge about an applicant. The use of resumes, for example, has a long history in which the resume document itself has been constructed as a reasonable proxy on which to base a hiring manager's judgements about an applicant that need to be examined in their historical context. Similarly, hiring ADSs that analyze tone of voice to discern personality characteristics have their epistemological roots in psychological profiles of discrete "personality types" and physiognomic approaches that [spuriously] link biological components of facial expression or vocalization to personality.[10,38]

Table 1: Examples of the type of information and ways of obtaining information for each element of the Socio-Technical Matrix.

| Element | Information | Questions and Method |
| --- | --- | --- |
| Hiring ADS | Name of hiring ADS | Question: What is the name of the hiring ADS? <br> Method: Identify from sales copy |
| Funnel Stage | Select from Bogen and Reike 2018 | Question: At what stage does this company's hiring ADS operate? <br> Method: Identify from sales copy and align with funnel list |
| Goal | Narrative description | Question: What is the hiring ADS intended to be used for? <br> Method: Identify from sales copy, interview developers and hiring managers who operate the hiring ADS |
| Data | Inventory of data types, datasets, benchmarking datasets | Question: What data, and what types of data, are used in training, testing, and operating the hiring ADS? <br> Method: Interview developers and hiring managers who operate the hiring ADS, inspect data directly |
| Function | Narrative description, machine learning models, metadata about models | Question: How does the hiring ADSwork? What is it optimizing for? <br> Method: Interview developers and hiring managers who operate the hiring ADS, inspect models, metadata, and product directly |
| Assumption | Narrative description | Question: Why is the hiring ADS useful? What is the assumed relationship between data about an applicant and the goals of the hiring manager? How does the hiring ADS inform the hiring process? <br> Method: Interview developers and hiring managers who operate the hiring ADS |
| Epistemological Roots | Narrative description | Question: Where do the assumptions made by the hiring ADS come from? What is their intellectual lineage? What are the critiques of this lineage? <br> Method: Archival research, interview developers and hiring managers who operate the hiring ADS, ethnographic study of hiring managers and developers |

## 3.2 Using the Matrix

In this section, we demonstrate how the matrix can be used (see Table 2). We completed the matrix ourselves using publicly available information about the product, as well as archival and historical research on common claims made by hiring tools. The landscape of automated tools used in the context of hiring is vast and emerging. As discussed above, most companies do not use just one hiring ADS, but combine various ADSs at various stages of the talent scouting and hiring process. Therefore, to demonstrate the use of the matrix in this short paper, we focus on **"screening"**, the second stage of the "hiring funnel".[31] This is the stage where candidates are assessed as to whether they are a good match for a job description. This assessment can be based on a myriad of aspects, but in analyzing the narrative description of several ADSs on the market, we identified several goals of available screening ADSs used to predict job fit in hiring: experience, skill, ability, and personality. An individual auditor would discern the specific goal(s) of the ADS as they work through the socio-technical matrix described above.

**Experience assessment** is the most basic form of assessment used in hiring and often focuses on using an analysis of education, previous positions, and years of experience as a proxy for job fit and future job performance. A standard way in which experience assessments happen is via parsing a resume.

**Skill assessment** is a form of standardized testing that sets out to measure a candidate's knowledge and skills that are needed for a particular role. For example, a very common skill assessment for programmers are so-called "coding challenges" whereby applicants are presented with typical programming challenges and have to solve live in a job interview.

**Ability assessment** typically refers to cognitive abilities tests. They are different from skill assessments, because they do not assess a skill that is learned, but are based on the assumption that there are latent mental abilities, such as abstract thinking, attention to detail, aptitude for understanding complex concepts, or adaptability to change, that not readily discernible from a a resume, cover letter, or interview.

**Personality assessments** set out to determine personality traits in an individual, such as introversion or extroversion. Aside from the infamous Myers-Briggs Type Indicator, a popular taxonomy is the OCEAN model that models the "Big 5 Personality Traits": openness to experience, conscientiousness, extraversion, agreeableness, and neuroticism.[39] In psychology, these personality traits are assumed to be stable,[40] and in hiring they are therefore thought to be predictive of on-the-job success. Personality tests have a long history in corporate management,[41] and automating them as part of screening assessments can be seen as falling well into the general shift towards the automation of general managerial decision-making.[42]

To show how the matrix can serve as a way to unpack how ADS construct experience, skill, ability, and personality, we offer examples from a selection of hiring ADSs marketed by several companies.

Table 2: An example of a completed Socio-Technical Matrix containing publicly-available information for several commercial hiring ADSs.

| **Hiring ADS** | Hiretual | Codility | Pymetrics | Humantic |
| --- | --- | --- | --- | --- |

| Funnel Stage | Screening | Screening | Screening | Screening |
| --- | --- | --- | --- | --- |
| **Goal** | Experience | Skill | Ability | Personality |
| **Data** | Resume<br>Professional profiles<br>Social media profiles<br>Proprietary database | Coding test exercises | Gameplay scores from applicants and workers | Resume<br>LinkedIn profile<br>Twitter profile |
| **Function** | Use profiling for job matching | Use test performance for screening candidates in / out | Use gameplay performance for screening candidates in / out | Use personality profiling for job matching |
| **Assumption** | Professional and social profile can be matched to job fit | Code test performance is a predictor of job skills | Gameplay is a predictor of job success | Personality is a good predictor for job fit |
| **Epistemological Roots** | Social Network Theory: The idea that who you are connected with reveals your identity. | Vocational Aptitude Testing: The idea that test scores predict ability. | Eugenics: The idea that intelligence and ability are innate and can be revealed through testing.[3] | Personality Types: The idea that personality is stable over time, and a predictor of of performance. |

## 4. Cues for auditability

The matrix can serve as a tool for developing new avenues for socio-technical work on the auditability of algorithms, beyond the already-existing dimensions along which ADSs are currently being audited. It unpacks how the hiring ADS conceptually constructs what it is supposed to measure and rank, for example experience, skill, ability, and personality.[43] Social research into the scientific and narrative roots of these frameworks can then help identify *how* theories underlying these measures are translated into constructs which are then operationalized in the hiring ADS.[44] For example, it allows us to trace how the idea that skill is relevant for assessing job fit became operationalized *as a construct* that underpins standardized assessments, such as standardized tests used in "technical interviews" (see section above). This way of mapping the construct genealogies that underlie assessments forms a bridge from qualitative investigation to technical inspection. It focuses attention on interrogating the *construct* as the basis for assessing *validity* without losing sight of the constructs' epistemological roots.

**Validity** generally refers to the extent that a statistical tool measures what it is supposed to measure.[5] Of particular relevance here is *construct validity*, which is "the extent to which the measure 'behaves' in a way consistent with theoretical hypotheses and represents how well scores on the instrument are indicative of the theoretical construct".[45]

---

[3] This idea has been well-debunked in the social sciences, which posit a critique that rests on such abilities as being socially constructed (Hacking 1986).

In other words, by using the matrix to clearly articulate what the goals of a particular ADS are *based on the construct that underpins it*, the technical system can be audited for validity (i.e. as to whether it achieves that goal equitably and reliably, or even at all). What is important is that the matrix, and the holistic and interdisciplinary approach it promotes, intervenes where a purely technical audit would end up inadvertently promoting a construct that is inherently problematic (e.g. by *exclusively* focusing on assessing construct validity and assessing if a tool "works as intended", rather than contextualizing the construct with the tool's intentions and claims to knowledge, and their histories). In other words, using the matrix to inform an audit prevents it from taking the construct at face value while *still* assessing its validity.

Personality testing as it is operationalized within hiring ADS provides an excellent case study. Here, we may use the matrix to identify function, assumption, and epistemological roots in order to identify the construct that underpins (automated) personality testing: personality as relatively stable, i.e. the idea that personality traits change little over time.[46] In other words, because some psychology literature assumes that personality is stable over time, although there is mounting disagreement with that,[47] we can assume that personality *as it is assessed via an ADS* should also be stable across systems, situations, and changing input factors (such as file type). Whether or not that is the case can be examined in an external audit using the same sample of individuals and their input data (such as CVs) to reveal instability in prediction across trials. Rhea et al. (forthcoming) have conducted such an audit of a selection of hiring tools and have demonstrated that stability is an important and accessible metric for external auditors seeking to inspect hiring ADS, using these inherent assumptions as baselines for audit design.

The matrix offered here for designing and running socio-technical audits can be used for assessing construct validity of many kinds of aspects that different ADS measure, because constructs carry the implicit claim that they are stable. This is the case for aspects that are measured by assessing candidates, but also beyond hiring domains. "Risk" is another such relevant construct that can be made auditable through this matrix, as operationalized in automated risk scoring systems that are used in the criminal justice system,[48] the public sector,[17] the healthcare sectors,[49] or the insurance industry.[50,51] For example, *person*-based risk scores that are used in predictive policing[48] should remain stable if, for example, the ZIP code associated with individual persons changes.

Using the matrix, the audit can then come full circle: audit results can and should be re-contextualized with function, assumption/s and epistemological roots of that tool to potentially reveal discriminatory effects which the creators of the tool would classify as "unintentional", and which can point toward the construct itself as being problematic or discriminatory by design, or alternately being sound and necessary for for making a hiring decision as required, in the U.S. by the Civil Rights Act of 1964[52] and clarified by the Equal

Employment Opportunity Commission.[4] By helping to identify the constructs that ADS, and hiring ADS specifically, claim to measure, the matrix can serve as a basis for more solid validity and reliability assessments that do not end up promoting discriminatory concepts or pseudoscience.

## 5. Toward robust audit frameworks

With the matrix as a starting point, we can better appreciate the complexity of auditing or analyzing hiring ADS and how this relates to regulatory audit mandates. While there is no industry-wide agreement on what audits ought to consist of, extant audits may ask about the Goal, Data, and Function of the ADS, but generally do not address issues of cross-model contamination (e.g. funnel stage), assumptions, and epistemological roots.

While the matrix is designed to be used by anyone wishing to conduct an *external* audit, and while audits may be conducted by anyone—developers, companies that purchase ADSs, public advocates, interested individuals—an ad hoc approach to conducting audits must eventually give way to a consensus about who ought to be tasked with auditing ADS. Following work on algorithmic impact assessment that calls for multidisciplinary perspectives on identifying algorithmic harms,[53] regulatory and legislative directives to conduct audits ought to require an interdisciplinary group of experts conduct audits, because of the range of expertise needed to productively interrogate the assumptions on which an ADS is based. In addition to legal and data science teams that are currently engaging in algorithmic audit work, we add our voices to the chorus of calls for the addition of social scientists, psychologists, and historians of science and technology to critically evaluate assumptions and epistemologies, and inform the audit process as a whole.[30,53–57]

## 6. Directions for Future Work

Auditing has also become a bit of a catch-all phrase, and there is value to parsing out different types of audits based on purpose and audience. In fields such as healthcare and finance, where audits are the norm, audit functions can be divided into two audiences: internal and external. Internal auditors are employed by the company, and external auditors can be a regulatory agency or a third-party group. Third-party groups can be a private firm specializing in audits or a potential client that may use the hiring ADS and wants to conduct their own audit. Further, the private firm may be compensated by a potential hiring ADS client OR by the company itself.

Whether the audit body is internal or external has significant impacts on: accessibility to models and data, chronology (i.e., when an audit is conducted in the development of this model), ability to assess cross-model contamination, and incentives.

Internal audit bodies have less external credibility but better access. In general, internal audit bodies serve to ensure the system is compliant with existing laws and addresses reputational risks. Internal audit groups may have access to data, models, IP, and key employees, which can mean engaging in the earliest stages of development, working closely with developers, data scientists, and project leads at milestones, and mitigating harm before there is adverse impact. These individuals are incentivized to ensure the company performs well, which can put into question the viability of fundamentally addressing issues of false assumptions and flawed epistemologies. Internal audit can range dramatically as ADS have no norms or laws dictating audit work. It is rarely the case that internal audits serve as external validation due to conflicts of interest. Generally, the role of internal audit is for organizational purposes, i.e. for product reliability and performance, or to ensure the company avoids legal or reputational backlash once the product is launched.[58]

---

[4] https://www.eeoc.gov/prohibited-employment-policiespractices

External audit bodies have more credibility but less access. Even if the audit is paid for by the company, external auditors have pressure to provide quality audit services to retain a good reputation. Regulatory bodies also publish their audit frameworks to allow internal audit teams to ensure adherence, but this also allows for public accountability. However, these auditors are not often granted unconstrained access to data, models, IP, or employees, and, in the current regulatory vacuum, companies have a heavy hand in creating these constraints. External audit bodies, however, are better able to critically analyze fundamentals such as assumptions and epistemologies as well as cross-model contamination.[59] Ultimately, the efficacy of any audit or assessment process will be judged by the material effects the process has on the minimization or mitigation of negative impacts for society and the environment. The Socio-Technical Matrix proposed here is just that, a proposal, and future work should entail applying the Matrix in practice, and developing a research programme to evaluate its effectiveness at ameliorating harms.

In the matrix, we outline a series of questions that need to be answered to understand how hiring ADSs are intended to work and how they actually operate in practice. Several of those questions require ethnographic investigation into the contextual uses and understanding of these hiring ADSs.

Some aspects of this ethnographic investigation are already well-established, particularly for conducting ethnographic interviews,[60] undertaking workplace ethnography,[61,62] and merging ethnographic fieldwork with archival research.[63] But methodological innovation is called for in tailoring ethnographic interviews, archival research, and fieldwork to robust accountability processes. Additional work is also needed beyond the investigation of specific hiring ADSs to better understand how hiring managers, workers, applicants, and others within an organization interact with each other and with hiring tools. Who are the hiring managers using the hiring ADS? When are hiring ADSs used in the pipeline and by whom?

Not all hiring ADSs are used the same way by hiring managers, some might take hiring ADSs as suggestions for their own decision-making, others might implement a hiring ADS's outputs directly, and the ways a hiring ADS functions within a hiring managers workflow ought to be inspected as part of any audit or impact assessment. Building out an understanding of how hiring ADS are used in the workplace is a job for ethnography. Gaining ethnographic insight into how job applicants experience and ascribe meaning to hiring ADS is equally important. Future work can and must focus on this side of the social practice of hiring. This ethnographic insight, then, will be essential not just for designing impactful audits, but also for—eventually—creating less invasive and discriminatory hiring technologies, for example in collaboration with worker organizations and unions.

## 7. Conclusion

In this paper, we have suggested a systematic approach for developing socio-technical assessment for hiring ADS. We have developed a matrix that can serve as a research tool for identifying the concepts that the hiring ADS claim to measure and rank, as well as the assumptions rooted in pseudoscientific essentialized understandings of human nature and capability that they may be based on. We have argued that the matrix can serve as a basis for more solid validity and reliability assessments, as well as a basis for critically investigating emerging auditing standards and practices that currently fail to address issues around pseudo-scientific approaches that essentialize membership in protected categories and make specious inferences from job-applicants' appearances and behaviors.

**ACKNOWLEDGMENTS**

The authors wish to thank the NYU Center for Responsible AI, Tandon School of Engineering, New York University and Tübingen AI Center, University of Tübingen. This work was supported by the German Federal Ministry of Education and Research (BMBF): Tübingen AI Center, FKZ: 01IS18039A. Additionally, the authors wish to thank the organizers and attendees of the "This Seems to Work" Workshop at the 2020 ACM CHI Conference, particularly Lindsey Cameron, Angele Christin, Michael Ann DeVito, Tawanna R Dillahunt, Madeleine Elish, Mary Gray, Rida Qadri, Noopur Raval, Melissa Valentine, and Elizabeth Anne Watkins. The authors also wish to thank the anonymous reviews for their helpful comments throughout.

**AUTHOR CONTRIBUTIONS**

All authors contributed equally to the writing of this piece, Dr. Mona Sloane designed the Scio-Technical Matrix.

**DECLARATION OF INTERESTS**

Rumman Chowdhury is General Partner, Parity Responsible Innovation Fund and sits on the Board of Advisors, Center for Data Ethics and Innovation, UK.

Dr. Mona Sloane is a Senior Research Scientist at the NYU Center for Responsible AI, Faculty at NYU's Tandon School of Engineering, a Fellow with NYU's Institute for Public Knowledge (IPK) and The GovLab, the Director of the *This Is Not A Drill* at NYU's Tisch School of the Arts, Technology Editor at Public Books, and a Postdoctoral Researcher at the Tübingen AI Center at the University of Tübingen, Germany. She also is on the Advisory Board of the Carnegie

Council for Ethics in International Affairs, and advises Fellowships at Auschwitz for the Study of Professional Ethics (FASPE).